# Irreducible Frequent Patterns in Transactional Databases

Gennady P. Berman[1], Vyacheslav N. Gorshkov[1,2], Xidi Wang[3]

[1] Theoretical Division, T-13, MS-B213, Los Alamos National Laboratory, Los Alamos, NM, 87545
[2] Theoretical Division, the Center for Nonlinear Studies, MS-B213, Los Alamos National Laboratory, Los Alamos, NM, 87545
[3] Citigroup, Sao Paulo, Brasil, 04543 906

Irreducible frequent patters (IFPs) are introduced for transactional databases. An IFP is such a frequent pattern (FP), $x_1 x_2 ... x_n$, the probability of which , $p(x_1 x_2 ... x_n)$, cannot be represented as a product of the probabilities of two (or more) other FPs of the smaller lengths, e.g.: $p(x_1 x_2 ... x_n) \neq p(x_1 x_2 ... x_k) \cdot p(x_{k+1} x_{k+2} ... x_n)$. We have developed an algorithm for searching IFPs in transactional databases. We argue that IFPs represent useful tools for characterizing the transactional databases and may have important applications to bio-systems including the immune systems and for improving vaccination strategies. The effectiveness of the IFPs approach has been illustrated in application to a classification problem.

## 1. Introduction

The fact that past behaviors can be used to predict future behaviors is the foundation of modeling in the business of consumer finance. Transactional patterns captured in historical databases reveal useful information about the nature of the transactions, allowing one to forecast the occurrence of similar patterns in the future. However, rarely occurring patterns are accidental and statistically insignificant. Frequent Patterns (FPs) [1-11], with frequency of occurrence above a certain threshold are statistically meaningful and particularly useful for financial modeling such as Market Basket Analysis (MBA) [1-6] etc.

Statistical regressions and linear, logistic, or nonlinear techniques such as neural networks are commonly used to predict consumer behavior for the population of an entire portfolio, while FPs can be used to identify outlier behaviors or pockets of opportunities within a smaller transaction subspace of interest. In such a scenario, we are interested in robust rules with high degree of confidence [12]. On the other hand, useful rules often involve high level multiple variable interactions. It is very difficult task for conventional techniques to build models with both high degrees of confidence and a high level of interactions. This is where FPs come into play. MBA is widely used in retail businesses for arranging floor layout and targeted marketing campaigns [1-6] where the requirement of a high level of robustness and variable interactions are less stringent. In addition, credit card companies often need to identify very sophisticated fraudulent patterns out of a vast amount of transactions where robust fraud detection rules are necessary to identify the fraudulent patterns while not inconveniencing good customers. In practice however, the number of FPs is often overwhelmingly large and many FPs identify the same occurrences. For practical applications, one needs to minimize the number of FPs required while maintaining the same level of effectiveness in order to reduce the implementation overheads and increase interpretability of the rules used. FPs can also be applied in many areas other than consumer finances.

Artificial Immune Systems (AIS) [13] can be built, mimicking their biological counterparts, for a number of machine learning and classification tasks with practical applications ranging from network intrusion detections [14] to bioinformatics [13]. Much research has gone into ways of generating good sets of non-self detectors or antibodies. In analogy to biological systems, these antibodies need to maintain their effectiveness while not exhausting precious resources such as T-cells and B-cells.

The objective of this paper is to construct irreducible FPs and their applications to computing scoring functions.

**Frequent Patterns.** Suppose that we have a set of elements $X = \{x_1, x_2, x_3, ..., x_K\}$. Any sub-set of these elements, with an arbitrary number of elements (or "length"), $x_i, x_j, x_k, ..., x_m$, is a single transaction. A transactional database ($DB_0$) is a set of $N$ transactions. Usually, a $DB_0$ can be characterized by at least two parameters, $K$-the maximal length of a transaction, and $N$-the total number of transactions. In addition to these two parameters, another useful characteristic of the $DB_0$ can be introduced, namely, a "pattern" which is an arbitrary set of the elements. A pattern can have a variable length, from $1$ to $K$. One important source of information is the frequency of the elements $x_k$, $f(x_k) \equiv f_k$ (patterns of the length one). The number $f_k$ indicates how many times the element $x_k$ appears in all transactions in the $DB_0$. More detailed information about the $DB_0$ is represented by the frequencies of the patterns of lengths $2,3,...,K$. For example, the frequency $f_{nm}$ is the number of patterns $x_n x_m$ in $DB_0$; the same is true for the pattern $x_n x_m x_l$, and so on. For many applications, the patterns with the frequencies less than some critical value (or threshold) $\xi$ are often not be considered as a characteristic property of the $DB_0$. Those patterns for which the corresponding frequencies satisfy the condition $f \geq \xi$ are called frequent patterns (FPs) [1-11]. Usually, the FPs are considered as important characteristics of the $DB_0$.

**Irreducible Frequent Patterns.** We argue that for many applications not all FPs are important characteristics of the transactional database $DB_0$. We demonstrate this statement below using a simple example of FPs $x_n x_m$ in the $DB_0$ for given values $f_n$ and $f_m$. For this, we shall build an auxiliary set of "equivalent" databases $DB_{e,l} (1 \leq l \leq L$, where $L$ is a large number of equivalent databases) each of them with the same number of transactions, $N$. We define the probability of appearance of the element $x_n$ in any transaction of any equivalent database to be $p_n = f_n / N$, where $f_n$ is, as before, the frequency of element $x_n$ in the database $DB_0$. A similar relation for the element $x_m$ is $p_m = f_m / N$. If the elements $x_n$ and $x_m$ are statistically independent, then the probability of the pattern $x_n x_m$ in any transaction is $p_{nm}^{(e)} = p_n p_m$. In this case, the frequencies $f_{nm}^{(l)}$ (and also $f_n$ and $f_m$) of the pattern $x_n x_m$, which are obtained for the databases $BD_{e,l}$, are random numbers. The average number of frequencies $f_{nm}^{(l)}$ over all databases, $BD_{e,j}$, and for $L \to \infty$, is $\langle f_{nm}^{(e)} \rangle = \lim_{L \to \infty} \frac{1}{L} \sum_{l=1}^{L} f_{nm}^{(l)} = N p_n p_m$. When comparing this average value with the "real" frequency $f_{nm}$ in the initial database $DB_0$, two possibilities can occur:

1. Suppose that the frequency $f_{nm}$ of the FP $x_n x_m$ in the database $DB_0$ differs only by a small value $\varepsilon$ from the average value $\langle f_{nm}^{(e)} \rangle$, namely $\left| f_{nm} - \langle f_{nm}^{(e)} \rangle \right| \leq \varepsilon$, (where the $\varepsilon$ is defined as the width of the probability distribution $P(f_{nm}^{(e)})$, see below). Then, we argue that the FP, $x_n x_m$, is not an important characteristic statistical property of the $DB_0$.

2. In the opposite case, $\left| f_{nm} - \langle f_{nm}^{(e)} \rangle \right| > \varepsilon$, the elements $x_n$ and $x_m$ are statistically dependent. In this case, the FP $x_n x_m$ is a new important statistical characteristic of the database $DB_0$. A measure of this statistical characteristic is the ratio $f_{nm} / \langle f_{nm}^{(e)} \rangle$.

If the elements $x_n$ and $x_m$ are statistically dependent, one can introduce a correlation coefficient between them, $c_{nm} \neq 1$. Suppose that each element $x_j$ takes only one of two values, $x_j = 1$ or $x_j = 0 \equiv \bar{x}_j$. We define the conditional probability, $p(x_m = 1 | x_n = 1) = p(x_m | x_n)$, of the element $x_m$, when the element $x_n$ already exists in the transaction ($x_n = 1$) as

$$p(x_m | x_n) = c_{nm} p_m, \qquad (1)$$

where the correlation coefficient can be represented in the form $c_{nm} = p(x_n | x_m) / p_n$.
On the other hand, $p(x_m | \bar{x}_n)(1 - p_n) + p(x_m | x_n) p_n = p_m$. Then,

$$p(x_m | \bar{x}_n) = p_m (1 - c_{nm} p_n) / (1 - p_n). \qquad (2)$$

When $c_{nm} = 1$, then $p(x_m | x_n) = p(x_m | \bar{x}_n) = p_m$, and the elements $x_n$ and $x_m$ are statistically independent. If $c_{nm} > 1$, then the elements $x_n$ and $x_m$ "attract" one each other. In other words, if the element $x_n$ exists in the transaction, then the probability for the element $x_m$ to occur in the same transaction is larger than $p_m$. In the opposite case, $c_{nm} < 1$, the elements $x_n$ and $x_m$ "repel" each other. Note that the correlation coefficient $c_{nm}$ can be represented in the form

$$c_{nm} = f_{nm} / \langle f_{nm}^{(e)} \rangle. \qquad (3)$$

One can see that the "interaction" between the elements in the database $DB_0$ is an important characteristic of its statistical properties. This "interaction" determines the properties of the FPs. From the total set of FPs, which can be quite large, one should choose only those FPs that have correlations among the elements of the database $DB_0$. The algorithm developed in our previous paper [15] allows one to search for all FPs in the database $DB_0$. As will be demonstrated below, this algorithm also allows one to select rather efficiently the IFPs. The required time for searching of IFPs is significantly smaller than that required to search all FPs.

## 2. Selection of IFPs

The procedure for searching for IFPs can be separated into two steps. 1) In the process of generating FPs we shall select those FPs that could be candidates for IFPs. The set of these FPs we shall denote as $G$. 2) We analyze a set $G$ and remove from it reducible FPs, keeping only IFPs.

Consider first the step 1. The selection of the manifold $G$ in $DB_0$ can be done by using the algorithm developed in [15]. Suppose that in $DB_0$ on the level $k$ of the main algorithm presented in [15] we obtained a FP $u$ of the length (the number of elements) $k$ with the frequency $f(u)$. The structure of this FP is $u \equiv x_n x_m ... x_j x_p$. In addition, the value $f(u)$ is known. Suppose that one needs to analyze on belonging to the set of IFPs a FP that is obtained by adding to the FP $u$ the element $x$ with the frequency $f(x)$ in the $DB_0$. As a result, one obtains a FP $v$ of the length ($k+1$) with the frequency $f(v)$. (For simplicity, we omit all indices in the expressions for $u$ and $x$.) Now, we shall build a set of "equivalent" databases using only two elements, $u$ and $x$. Suppose that in each "transaction" these elements appear randomly with the probabilities $p(u) = f(u)/N$ and $p(x) = f(x)/N$. Let's assume that the elements $u$ and $x$ are statistically independent. Represent each transaction by a two-dimensional vector on the plane $ux$, with the components taking two values 0 or 1. Then, the vectors $(0,1)$, $(1,0)$ represent transactions which include only one of the elements, $x$ or $u$. The vector $(1,1)$ represents a transaction which includes simultaneously both elements. The equivalent databases are represented by the matrices, each of them with 2 lines and $N$ columns. (See Table 1, where one such matrix is shown.)

| $i$ | 1 | 2 | … | N-7 | N-6 | N-5 | N-4 | N-3 | N-2 | N-1 | N |
|---|---|---|---|---|---|---|---|---|---|---|---|
| $x^{(l)}$ | 0 | 1 | … | 0 | 0 | 0 | 1 | 1 | 0 | 1 | 1 |
| $u^{(l)}$ | 1 | 0 | … | 1 | 1 | 1 | 0 | 1 | 0 | 0 | 1 |

Table 1: The equivalent database with the number $l$ ($l$=1,2,…,$L$, where $L$ is the number of the equivalent databases); $i$ is the number of transaction.

If one assumes that the element $x$ does not correlate with the FP $u$, then the average frequency of the FP $v$ in the equivalent databases is

$$\left\langle f^{(e)}(v) \right\rangle = N \frac{f(u)}{N} \frac{f(x)}{N} \equiv N p(u) p(x) \equiv f^0(v),$$

where $\langle ... \rangle$ indicates, as before, averaging over $l$.

The "real" frequency $f(v)$ in the database $DB_0$ can be different from $f^0(v)$. The information on possible correlations between $u$ and $x$ can be obtained by estimating the expression $\Delta = |f(v) - f^0(v)|$. If this value is larger than some threshold $\varepsilon$ (which will be determined below), than the elements $u$ and $x$ are correlated.

**Estimate of the threshold $\varepsilon$.** If the elements $x$ and $u$ are statistically independent, then the frequency $f^{(l)}(v)$, at given probabilities $p(u)$ and $p(x)$, is a random number. This frequency can be represented as

$$f^{(l)}(v) = \sum_{i=1}^{N} u_i^l x_i^l, \tag{4}$$

where $u_i^l, x_i^l$ take the values 0 or 1, in correspondence with the matrix in the Table1. When the number of transactions $N$ is large, the random number $f^{(l)}(v)$ has the normal distribution with the probability density which has the mean square deviation $\sigma$:

$$\sigma^2 = \left\langle \left(\sum_{i=1}^{N} u_i^l x_i^l\right)^2 \right\rangle - \left(f^0(v)\right)^2. \tag{5}$$

To calculate the dispersion $\sigma$, the following expression can be used

$$\left\langle \left(\sum_{i=1}^{N} u_i^l x_i^l\right)^2 \right\rangle = \left\langle \sum_{i=1}^{N} (u_i^l x_i^l)^2 \right\rangle + \left\langle \sum_{i \neq j} (u_i^l x_i^l)(u_j^l x_j^l) \right\rangle = Np(u)p(x) + (N^2 - N)[p(u)p(x)]^2. \tag{6}$$

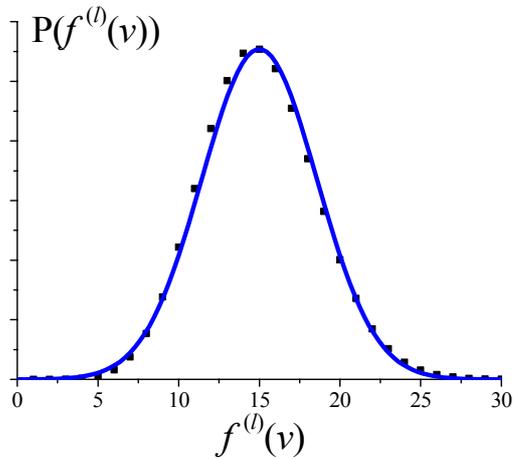

Fig. 1: A solid curve represents a distribution (8). The points are the result of numerical simulations for $L = 3 \cdot 10^6$ equivalent databases randomly generated with $N = 100$, $q_{ux} = 0.15$; $\sigma \approx 3.6$.

Taking into account that $f^0(v) = Nq_{ux}$, where $q_{ux} = p(u)p(x)$, we derive from (6)

$$\sigma^2 = Nq_{ux}(1 - q_{ux}). \tag{7}$$

Then, the probability distribution $P\left[f^{(l)}(v)\right]$ of the random number $f^{(l)}(v)$ cab be represented in the form

$$P\left[f^{(l)}(v)\right] = \frac{1}{\sigma\sqrt{2\pi}} \exp\left[-\left(f^{(l)}(v) - Nq_{ux}\right)/(2\sigma^2)\right] \tag{8}$$

The validity of the expression (8) is demonstrated in Fig. 1. A solid curve represents the function (8) for $N = 100$, $q_{ux} = 0.15$. The squares are the results of numerical simulations. In this case, the equivalent databases were randomly generated with given parameters, and $L = 3 \cdot 10^6$. Usually, the deviation of $f^{(l)}(v)$ from the average value $f^0(v)$ does not exceed $3\sigma$ (a known fact for the normal distribution). If the real value of the frequency $f(v)$ in the database $DB_0$ exceeds $3\sigma$

$$|f(v) - Nq_{ux}| > 3\sigma, \tag{9}$$

then one can argue that the element $x$ correlates with the FP $u$ in the database $DB_0$. In this case, the FP $v$ must be included in the manifold $G$. The inequality (9) can be presented in the form

$$\left|\frac{f(v)}{Nq_{ux}} - 1\right| > \gamma \sqrt{\frac{1-q_{ux}}{Nq_{ux}}}, \tag{10}$$

or

$$|w_{u;x}| \equiv \left|\frac{f(v)}{Nq_{ux}} - 1\right| \bigg/ \gamma \sqrt{\frac{1-q_{ux}}{Nq_{ux}}} > 1, \tag{11}$$

where the value $w_{u;x}$ will be called the level of correlation of elements $u$ and $x$. The value of parameter $\gamma = 3$ establishes a threshold for selection of IFPs. In this case, if $w_{u;x} \leq 1$, than the elements $u$ and $x$ will be considered as statistically independent.

Note, that in some cases the coefficient $\gamma$ can be chosen $\gamma < 3$ (for example, $\gamma \approx 2$). Remind that non-correlating elements lead to the normal distribution (8) for the frequency distribution function of the corresponding FPs. For $\gamma = 3$ the criterion (10) drops out practically 100 % of non-correlating FPs. For $\gamma = 2$ the criterion (10) drops out approximately 95% of these patters.

The described above procedure for selection of IFPs which belong to the manifold $G$ is easy to implement in the algorithm [15]. The computer time will not practically increase in comparison with the time required for searching of all FPs. However, the solution of the problem for selecting of IFPs will require additional steps. Namely, the manifold $G$ will still include many FPs (the number of which could, in principal, significantly exceeds the number of transactions in the database) which do not belong to the IFPs[*]. The reason for this is that criteria (10) or (11) do not guarantee that obtained FPs cannot be reduced to the IFPs of smaller lengths. The problem we are facing now is the following—how to minimize the number of reducible FPs in G? Below we discuss the method for solving this problem. We also tested our approach on artificially generated databases.

---

[*] The reason why some of the FPs which satisfy the criterion (10) could be reducible, is illustrated below on concrete examples.

**Generation of an artificial database $DB_0$ with correlated elements.** In this sub-section we describe the method which was used for generating artificial databases $DB_0$ with correlated elements. Every transaction in the database is represented by a set of $M$ variables $Y_1, Y_2, Y_3, ..., Y_M$. In our numerical simulations, the number of variables, $M$, was equal to 30. Each variable, $Y_i$, takes only one value from the set of the values $y_{i1}, y_{i2}, ..., y_{im_i}$. We also assume that all $m_i$ ($i=1,2,...,M$) are the same for all variables, $Y_i$, and are equal to 3 ($m_i = 3$). The database $DB_0$ can be represented by a rectangular matrix which is filled with ones and zeros. (See Fig. 2.) Each column in the matrix represents a single transaction, the ones indicate the values $y_{ij}$ ($i=1,...,30$; $j=1,2,3$) which are included in a given transaction. In each row, the ones indicate transactions containing a given value, $y_{ij}$. Below, instead of the values $y_{ij}$ we use the values (elements) $x_k$ ($k = 3(i-1) + j$); the total number of which is $K = 90$ ($k=1,2,...,90$).

|     |          |           |   |   |   |   |   |   |   |
|-----|----------|-----------|---|---|---|---|---|---|---|
|     | $y_{M3}$ | $x_K$     | 0 | 1 |   |   |   |   | 1 |
| $Y_M$ | $y_{M2}$ | $x_{K-1}$ | 0 | 0 |   |   |   |   | 0 |
|     | $y_{M1}$ | $x_{K-2}$ | 1 | 0 |   |   |   |   | 0 |
|     |          |           | • | • |   |   |   |   | • |
| •   |          |           | • | • |   |   |   |   | • |
|     |          |           | • | • |   |   |   |   | • |
|     | $y_{j3}$ | $x_{l+2}$ | 0 | 0 |   |   |   |   | 0 |
| $Y_j$ | $y_{j2}$ | $x_{l+1}$ | 1 | 0 |   |   |   |   | 0 |
|     | $y_{j1}$ | $x_l$     | 0 | 1 |   |   |   |   | 1 |
|     |          |           | • | • |   |   |   |   | • |
| •   |          |           | • | • |   |   |   |   | • |
|     |          |           | • | • |   |   |   |   | • |
|     | $y_{23}$ | $x_6$     | 0 | 0 |   | Interaction |   |   | 0 |
| $Y_2$ | $y_{22}$ | $x_5$     | 0 | 1 |   |   |   |   | 1 |
|     | $y_{21}$ | $x_4$     | 1 | 0 |   |   |   |   | 0 |
|     | $y_{13}$ | $x_3$     | 1 | 0 |   |   |   |   | 1 |
| $Y_1$ | $y_{12}$ | $x_2$     | 0 | 0 |   |   |   |   | 0 |
|     | $y_{11}$ | $x_1$     | 0 | 1 |   |   |   |   | 0 |
| Number of transaction | | | 1 | 2 | • | • | • |   | N |

Fig. 2. Illustration of the technique for generating of the artificial database $DB_0$.

Suppose that all rows of the matrix in Fig. 2 for the variables $Y_1, Y_2, ..., Y_{i-1}$ are filled up. For the element $x_l$ ($l = 3(i-1)+1$), which represents a single value $y_{i1}$ of the variable $Y_i$, we definite the probability, $p_l$, of appearance of the element $x_l$ in the transactions in the database $DB_0$:

$$p_l = p_{max} \times \zeta, \qquad (12)$$

where $\zeta$ is the random number uniformly distributed within the interval from 0 to 1[**], and $p_{max}$ is a parameter for the generated matrix (database). The probability $p_l$ will be used below as the "average" probability of appearance of the element $x_l$ in the row $l$. This probability does not include the information about possible correlations between the element $x_l$ and other elements $x_k$ in the database with $k<l$. To introduce these correlations we use the following procedure. Among the elements $x_{k<l}$, we define the element $x_k$ with which the element $x_l$ interacts. The number $k$ is defined from the expression: $k = [(l-1)\zeta + 1]_{int}$, where $[z]_{int}$ is the integer of $z$. In Fig. 2 the element $x_l$ correlates with the element $x_3$. The corresponding correlation coefficient is defined as

$$c_{kl} = 1 + \theta \times \zeta, \qquad (13)$$

where $\theta$ is some parameter (see below for details) which defines the maximum of correlation, $1 \leq c_{kl} \leq 1+\theta$. Now we shall fill up the row with the number $l$ in the database $DB_0$ using the following rule. If in a transaction with the number $n$ (the column $n$) $x_k = 1$, the conditional probability of appearance of the element $x_l$ ($x_l = 1$) in the same transaction is equal to

$$p(x_l = 1 | x_k = 1) \equiv p_{11} = c_{kl} p_l . \qquad (14)$$

Then, having drawn a new random number $\zeta$, we define the element $x_l$:

$$x_l = \begin{cases} 1, & \zeta \leq p_{11} \\ 0, & \zeta > p_{11} \end{cases}. \qquad (15)$$

In the case when $x_k = 0$, we use the conditional probability $p(x_l = 1 | x_k = 0) \equiv p_{01} = p_l(1 - c_{kl} p_k)/(1 - p_k)$ (see Eq. (2)):

---

[**] Below we shall use the random number generator $\zeta$ not only for generating $p_l$, but also for many other purposes. To simplify the expressions, each time when a random number generator $\zeta$ occurs, it is assumed that a new random number was generated.

$$x_l = \begin{cases} 1, & \zeta \le p_{01} \\ 0, & \zeta > p_{01} \end{cases}. \qquad (16)$$

After the row with the number $l$ is filled up, we build a conditional probability $p(x_{l+1}=1|x_l=0) = p_{max} \times \zeta$ (if $x_l = 1$, then $x_{l+1}$ is definitely equal to 0 since the variable $Y_j$ can take only a single value 1 from the set $y_{j1}, y_{j2}, y_{j3}$). Using this probability we fill up the row with the number $l+1$ by using the equation similar to Eq. (15). To complete the operations with the variable $Y_i$, we have to determine $x_{l+2}$ for all transactions (columns) in the database $DB_0$, by using a simple rule:

$$x_{l+2} = (1-x_l)(1-x_{l+1}). \qquad (17)$$

When the initial stage of the procedure is carried out (i.e. the rows of the matrix for the variable $Y_1$ are filled up), we use the technique presented above, but the element $x_1$ is supposed to be a statistically independent of $x_k$ with $k>1$. In each column of the matrix the element $x_1$ is defined according to the rule:

$$p_1 = p_{max} \times \zeta, \quad x_1 = \begin{cases} 1, & \zeta \le p_1 \\ 0, & \zeta > p_1 \end{cases}. \qquad (18)$$

As a result, in the database $DB_0$, generated this way, the multiple correlations exist among the elements, and the properties of these correlations are generally unknown. Our next goal is to select IFP by building the manifold $G$ and by using the developed below technique. The specific parameters of the $DB_0$, $M = 30$ and $m = 3$, used in this paper for illustration, do not play a principal role, and the results can be applied for any $M$ and $m$.

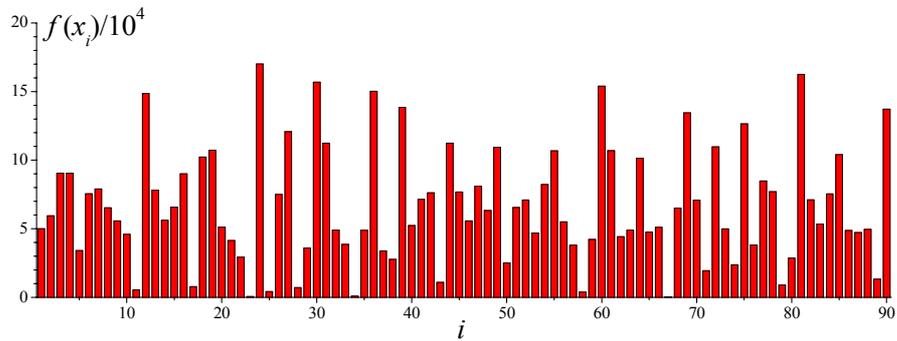

Fig. 3. The frequencies of the elements $x_i$ in the artificial database $DB_0$; $p_{max} = 0.6$, $N = 200000$.

**Building the manifold** $G$. Usually, the initial database $DB_0$ is a set of transactions that are generated according to some unknown rules (in many cases these rules do not exist). In such a database, the interaction between the elements is unknown and can be quite strong. To model this type of database, we generated the correlation coefficients which are bounded from below by the value 1.3:

$$c_{nm} = 1.3 + \theta \times \zeta, \quad \theta = 0.1. \tag{19}$$

The total number of correlation coefficients $c_{kl}$ between two elements $x_k$ and $x_l$, which were determined by the explicit method (see Eq. (13)), is 29. Besides, since the values $y_{i1}$, $y_{i2}$, $y_{i3}$ of the variable $Y_i$ are bound by the relation $y_{i1} + y_{i2} + y_{i3} = 1$, many uncontrolled correlations are expected to appear in the database $DB_0$, which contribute to the IFPs of different lengths. If the threshold for searching FPs is $\xi = N/50$ (2% of the total number $N$ of transactions, $N = 200000$), the number of FPs in the database $DB_0$ (with frequencies of the elements presented in Fig. 3) is approximately 4.5 millions.

The preliminary fast selection of the manifold $G$ (the "express" analysis) of the database $DB_0$ one can do as follows.

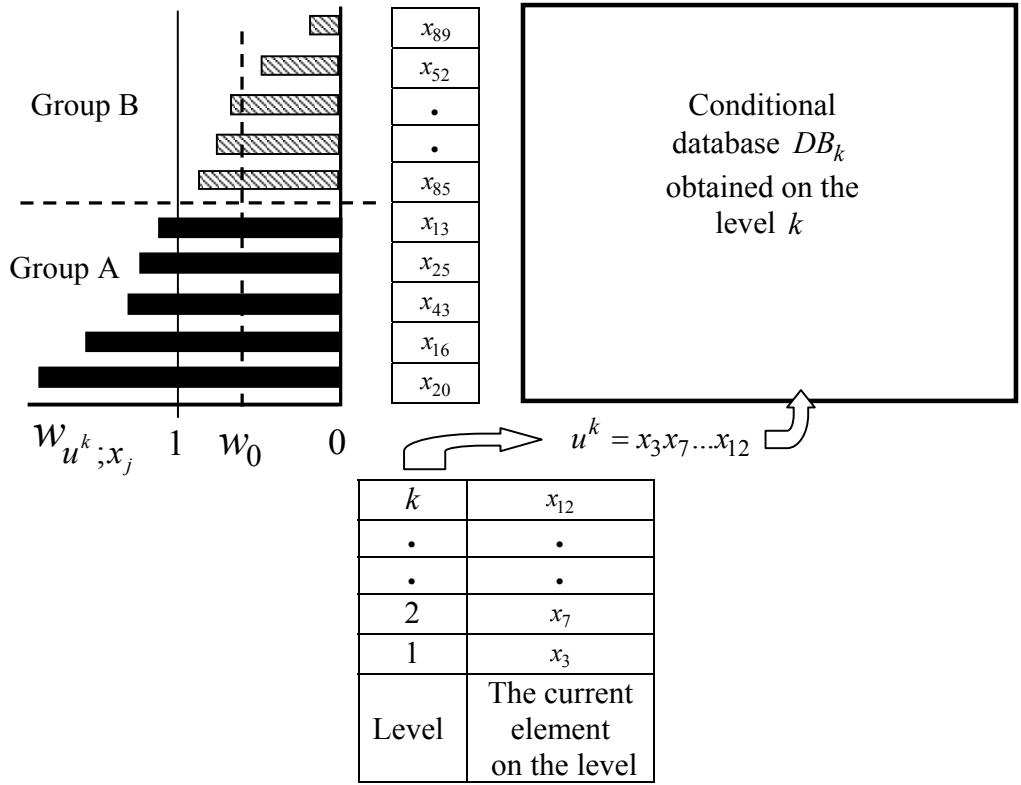

Fig. 4. The illustration of the calculation process on the level $k+1$. The combination of basic elements, starting from the level 1 through the level $k$; $u = x_3 x_7 ... x_{12}$ in this case, we shall call a "radical" for the level $(k+1)$ and denote it as $u^k$.

In the conditional database $DB_k$, which was built on the level $k$ of the main algorithm [15] (and is processed on the level $(k+1)$), represent the elements in the order of decreasing of the level of correlations $w_{u^k;x_j}$ (see Eq. (11) and Fig. 4). Then, the conditional databases, $DB_{k+1}$, generated on the level $(k+1)$ (for subsequent processing them on the level $(k+2)$), we will build only for the group of elements $A$, for which $w_{u;x_j} > w_0 = 1$. It is clear that FPs generated on the level $(k+1)$, $v_j = u^k x_j$, $x_j \in A$, should be considered as belonging to the manifold $G$ because the elements $x_j$ evidently correlate with $u^k$. This procedure is performed on all levels of the main algorithm [15]. If on some level, $l$, a conditional database $DB_l$ will be generated with an empty manifold $A$ for the radical $u^l$, than the generation of the elements $v_j = u^l x_j$ on the level $(l+1)$ for the manifold $G$ is not performed. In this case, one returns to the cycle on the level $k$ with generation of a new radical $u^l$ and a new database $DB_l$. A maximum number of enclosed levels (cycles), к, which is a parameter of the problem, determines the maximal length of the IFPs generated by using this approach.

We realized a described above procedure (for к=10) for the artificial database (Fig. 4 demonstrates a part of the procedure), for the parameter $\gamma = 2$. In the manifold $G$ 3009 FPs were selected with the maximal length $l_{max} = 7$ (in spite of the FPs with larger $l$ satisfied the condition $f \geq \xi = N/50$ for their frequency, $f$, in $DB_0$, they were reducible). The time needed for generation of the manifold $G$ is two order of magnitude smaller than the time required for searching all FPs. In the process of analysis of the manifold $G$ (see below) 1803 "real" IFPs were selected. A distribution of the number of IFPs, $U(l)$, with a given length, $l$, is shown in Fig. 5. However, a part of IFPs is excluded from consideration under this procedure (Fig. 5). We now describe the structure of the lost IFPs, in order to develop a strategy for their searching without a significant increase of the searching time. If one sets a threshold $w_0 = 0$ and builds conditional databases on the level $(k+1)$ for all elements $x_j$, than a generation of the following IFPs on the level $(k+2)$ becomes possible (Fig. 6).

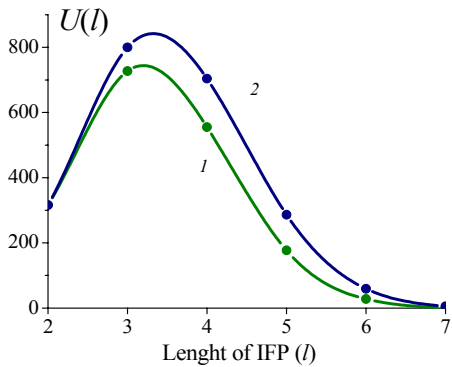

Fig. 5. Distribution of the number of IFPs of a given length for different values of the threshold $w_0$; $w_0 = 1$ (curve 1) and $w_0 = 0$ (curve 2).

When in the criterion (10) the parameter $\gamma$ is chosen $\gamma = 2$, the total number of IFPs is 82 (except from the single elements $x_i$). We denote a particular IFP of the length $m$ as $F_m$ (a feature of the database $DB_0$), and by $^m F$ we denote a total number of $F_m$ in $DB_0$. For $N = 50000$ the results of simulations are the following: $^2F = 54$, $^3F = 27$, $^4F = 1$. If the number $N$ of transactions in the database $DB_0$ increases (while the probabilities $p_l$ and the correlation coefficients $c_{kl}$ are not changed), the total number of IFPs increases, too. (See Tab. 2.) One can see from Table 2 that the function $^2F(N)$ becomes saturated when $N$ increases. This phenomenon has a simple explanation. The mean square deviation $\sigma$ (in the probability distribution (8) for the frequency $f^{(l)}(x_i x_j)$ of independent elements $x_k$ and $x_l$) depends on $N$ as $\sqrt{N}$. Similarly, the mean square deviation s).

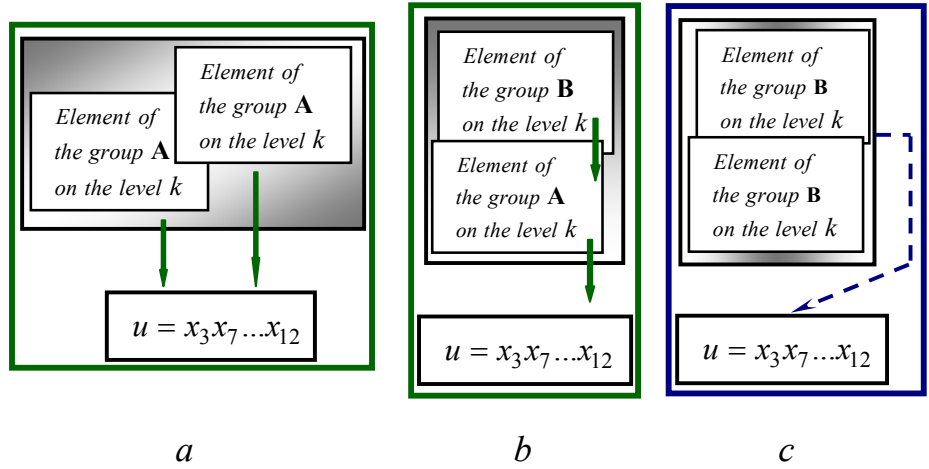

Fig. 6. Possible types of IFPs generated on the level $(k+2)$.

**Case** $a$: The most obvious case of generation of the element in $G$ is realized when the candidate in IFPs on the level $k+2$ is formed by adding to the radical $u$ two interacting with $u$ elements, $x_{j_1}^{(A)}$ and $x_{j_2}^{(A)}$, of the group $A$ in the database $DB_k$ (see Fig. 4).

**Case** $b$: A FP of the length $k+2$ is built due to the interaction between the element of the group $B$, $x_{j_3}^{(B)}$, with the element of the group $A$, $x_{j_1}^{(A)}$, or with the combination $ux_{j_1}^{(A)}$.

The candidates in the IFPs, which are generated in the cases $a$ and $b$, are extracted by the described above procedure for building of the manifold $G$, because on the level $k+1$ all conditional databases for the elements of the group $A$ are built (Fig. 4). During this procedure, on each step of the algorithm one adds (on the level $k+1$) to the radical $u$ of the level $k$ only the interacting with $u$ elements. Namely, under generation of candidates in IFPs only the "direct" interactions of the radical $u$ with the elements of the database $DB_k$ are taken into consideration.

**Case** $c$: This is an example of IFPs of a more complicated structure which could be "invisible" under the procedure described in the beginning of this sub-section (see also Fig. 4). Two elements of the group $B$, $x_{j_3}^{(B)}$ and $x_{j_4}^{(B)}$, do not interact separately with the radical $u$ (and, possibly, between themselves). However, the combination of these elements, $x_{j_3}^{(B)} x_{j_4}^{(B)}$, could interact with the radical $u$ (see Figs. 4 and 6). This is an example of "indirect" interaction or the second order interaction.

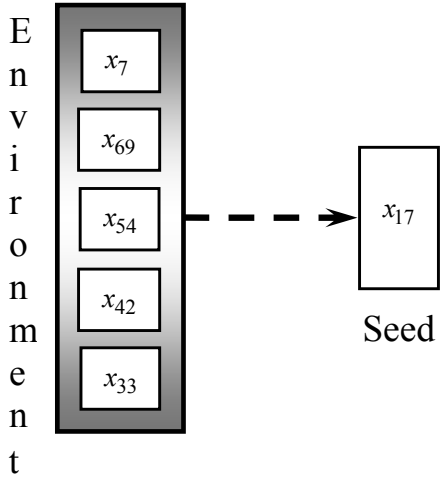

Fig. 7. The IFP, $(x_7 x_{69} x_{54} x_{42} x_{33}) x_{17}$, produced by the indirect interaction of the fifth order. The existence of such IFP in the database could lead to the existence of IFPs of lower order, for example, $(x_7 x_{54} x_{42}) x_{17}$.

In the database $DB_0$ the indirect interactions of a more higher order could exist. In Fig. 7, an example of IFP is demonstrated which is produced by the fifth order indirect interaction. The elements of the "environment" (see Fig. 7) could be statistically independent, and only a cornel ("seed" $x_{17}$) combines them in the IFP $(x_7 x_{69} x_{54} x_{42} x_{33}) x_{17}$. The difficulties of selection of such IFPs could be related to the fact that under the large value of the correlation coefficient, $c_{(7,69,57,42,33),\ 17}$, between the environment and the seed, the correlation coefficients of pair correlations (direct interactions) could be close to one. To demonstrate this, we present below an example.

Suppose that the elements of the environment are statistically independent, and their probabilities in the database $DB_0$ are given as follows: $p_7 = 0.3$, $p_{69} = 0.325$, $p_{54} = 0.35$, $p_{42} = 0.375$, $p_{33} = 0.4$. For the "seed" $p_{17} = 0.425$, and for the correlation coefficient $c_{(7,69,54,42,33),\ 17} = 1.4$. In this case, (after generation of the artificial database $DB_0$) the coefficients of pair correlations were calculated: $c_{7,17} = 1.0052$, $c_{7,69} = 1.0038$,… $c_{7,33} = 1.0022$ (the largest is the probability of the elements of the "environment", the smallest is the correlation coefficient with the "seed"). Because the coefficients of pair correlations are close to 1, to detect these direct interactions is a difficult task, especially if the number of transactions in the database $DB_0$ is rather small (see Eq. (10)). However, in the main algorithm, the interaction of the elements in the IFP, shown in Fig. 7, is tested at each level of the procedure. Suppose, that in the simplest case the radical $u$ on the level 1 is the "seed" $x_{17}$. For the threshold $w_0 = 1$ not all elements of the environment could belong to the group $A$ in the conditional database built for a given radical and processed on the level 2. (In particular, it is possible that all elements of the environment will appear in the group $B$).

In order to select IFPs with a high order indirect interactions we use a reduction of the threshold $w_0$ ($w_0 < 1$), which separates the elements in the conditional database for the radical $u$ (level $k$) for groups $A$ and $B$ (level $k+1$). Note, that this procedure only increases the number of FPs generated on the level $k+1$, but does not reduce the threshold of IFPs which belong to the manifold $G$ and which are defined by Eq. (10). The work of the algorithm in this case will be as follows. Assume that the interaction of the elements $x_7$ and $x_{17}$ (Fig. 7) will not be detected on the level 2 ($w_{17;7} < 1$). However, in the case $w_{7;17} > w_0$ we shall continue calculations with the radical $x_{17} x_7$. The correlation coefficient of the element $x_{69}$ with the radical $x_{17} x_7$ on the level 3 increases and equals to $c_{(17,7),69} = 1.0145$ (in comparison with $c_{7,17} = 1.0052$ on the level 2), which significantly exceeds the probability to detect the interaction on the level 3. Moreover, on the level 4, for the radical $x_{17}x_7x_{69}$ and the element $x_{54}$ the corresponding correlation coefficient is $c_{(17,7,69),54} = 1.04$, which is big enough for selection of this IFP. (For the level 5 $c_{(17,7,69,54),42} = 1.096$, and for the level 6 $c_{(17,7,69,54,42),33} = 1.20$). Thus, starting with some level $k$ we start to detect indirect interactions of the order $k-1$. However, in the databases with a large number of transactions, when the parameter $\sigma/N$ in the distribution (8) is small, presented in Fig. 7 IFP can be detected through the direct interactions of the environment with the "seed".

Note, that when the threshold $w_0$ decreases, the manifold $G$ includes not only the IFPs but also many reducible FPs. The reason why at small values of $w_0$ the reducible FPs appear in the manifold $G$ is the following. Suppose that we are searching all IFP with the maximal length κ=8. Suppose also that on the level 3 we select the IFP $x_{12}x_{73}x_{57}$. Then, on the levels 4 and 5 the statistically independent elements $x_{25}$ and $x_{44}$ are added. At the threshold $w_0 = 1$ we would not consider any candidates for IFPs which include these elements. However, at smaller threshold $w_0$ we should consider a FP of the length 5, $(x_{12}x_{73}x_{57})(x_{25}x_{44})$, as a part of a possible IFP with a high order interaction with the element, say, $x_{77}$ on the level 6, and with the length 6. However, suppose that on the level 6 the element $x_{77}$ strongly interacts only with a pair $(x_{12}x_{57})$. Then, the FP $(x_{12}x_{73}x_{57})(x_{25}x_{44})x_{77}$ will be included in $G$, but it is a reducible FP. Indeed, the probability of this FP can be represented as a product $p(x_{12}x_{73}x_{57}x_{77}) \cdot p(x_{25}) \cdot p(x_{44})$. In the worst scenario, between the IFP $(x_{12}x_{73}x_{57})$ and the element $x_{77}$ on the levels with numbers from 4 to 7 many non-interacting elements could be involved, which leads to appearance of many reducible FPs in $G$. That is why it is useful to introduce a parameter $R$, which limits the number of elements in the radical $u$ (see Figs. 4 and 6) which have the level of correlations $w$: $w_0 < w < 1$. For example, suppose that on the level 6 the radical appears

$$x_{28}x_{21}(x_{39})x_{83}(x_{98})(x_2)$$

(in parentheses are the elements which do not satisfy the criterion (10) in relation to the group of elements which are located on the left from them). If $R = 3$, than in the numerical procedure the transition to the level 7 is not performed (because in the initial radical three non-interacting elements already exist). At the same time, the algorithm continues to operate on the level 6. Namely, the algorithm tests the combinations of other elements in the conditional database $DB_5$ with the radical $x_{28}x_{21}(x_{39})x_{83}(x_{98})$ on their belonging to the IFPs. Thus, the parameter $R$ limits the order of registered indirect interactions by the value $(R-1)$. In other words, our algorithm detects only IFPs which represent a "seed" (generally with arbitrary number of elements) and with $(R-1)$ elements of environment. Note, that the parameter $R$ does not limit the length of the candidates in IFPs but only the order of the registered interactions.

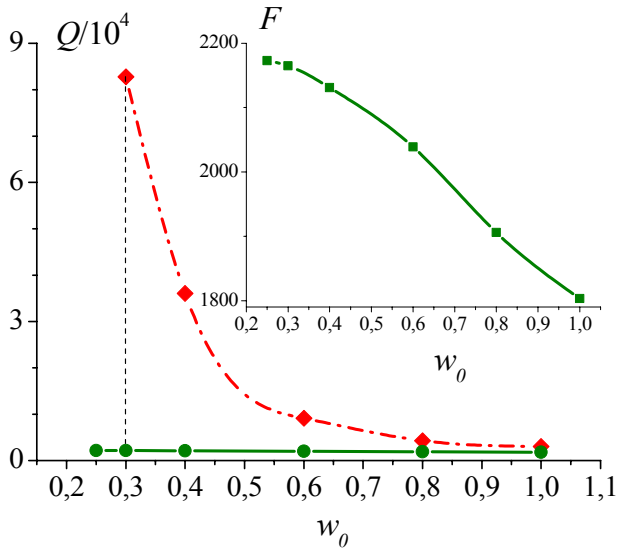

Fig. 8. The dependence of the number of FPs, $Q$, in the manifold $G$, and the number, $F$, of selected $IFP$, $F$, on the threshold, $w_0$.

Now we present the numerical data which illustrate our algorithm for different parameters (see Fig. 8). One can see that the "express-analysis" with $w_0 = 1$ and $R=1$ finds a major part of IFPs. The increase of the number $F$ of IFPs when $w_0$ decreases is caused by the indirect interactions. The maximum value of $F$, $F_{max}$, does not exceed 2200. This number is in agreement with the intuitive requirement: $F_{max} << N$ (the total number of characteristic features, $F_{max}$, of the database $DB_0$ should not exceed the total number of transactions).

The decrease of $w_0$ results in exponential increase of the number of FPs in $G$ (red curve in Fig. 8). If to introduce the parameter $R$, for example, $R = 3$, than (i) the number $Q$ in the manifold $G$ significantly (more than the order of magnitude) increases and (ii) the number $F$ of IFPs practically does not change (i.e., in this case, in the database $DB_0$ the indirect interactions of the order more than two do not exist). The time which requires to search IFPs, even for $w_0 = 0.25$ is al least one order of magnitude smaller than the time $T$ which is required to search all FPs in the database $DB_0$.

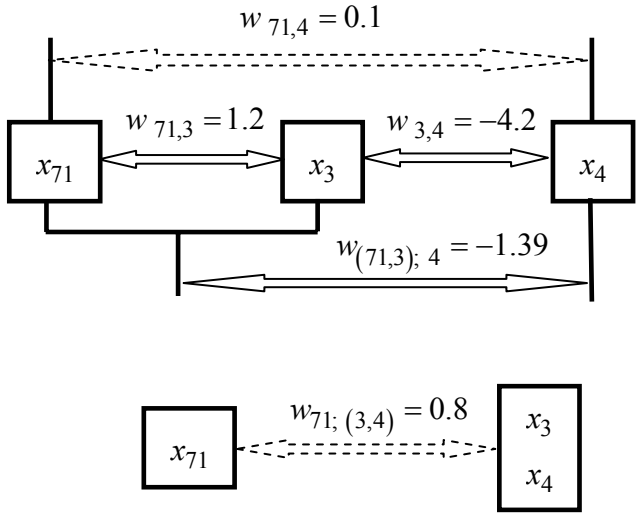

Fig. 9. A scheme which illustrates a decomposition of a FP on IFPs.

Note, that even at $w_0 = 1$ the reducible FPs appear in the manifold $G$. To illustrate this, we present a FP of the length 3, $x_{71}x_3x_4$, which was generated in the numerical simulations (Fig. 9; in the database $DB_0$ the elements were presented in the order of increasing of their frequency). On the level 2 the IFP $x_{71}x_3$ was generated with strong interaction, $w_{71,3} = 1.2$. The next element, $x_4$, actively interacts with both the element $x_3$ and with the pair $x_{71}x_3$ (interaction of $x_3$ with $x_{71}$ is absent). However, the analyzed FP $x_{71}x_3x_4$ cannot be referred to IFPs as it can be reduced to the non-interacting IFPs $x_{71}$ and $x_3x_4$ (Fig. 9; $w_{71;(3,4)} < 1$):

$$p(x_{71}x_3x_4) = p(x_{71})p(x_3x_4). \qquad (20)$$

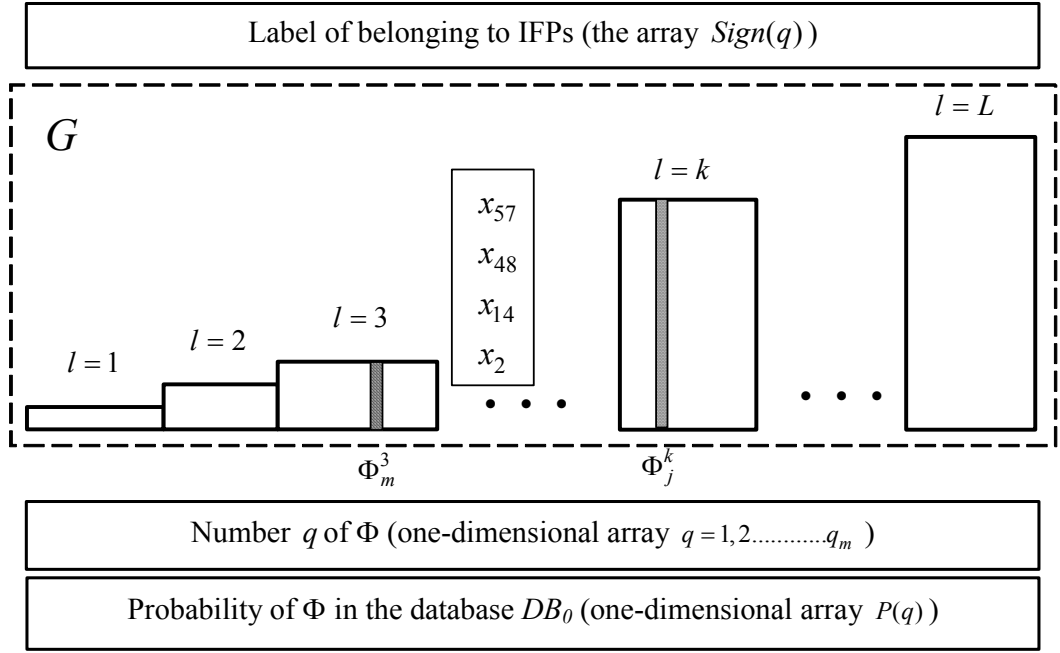

Fig. 10. The structure of the manifold $G$. The elements $\Phi$ of the same length $l$ are collected in the groups (in the insert an example is shown of such element of the length 4, $\Phi^4$, as a column of the matrix which represents a database $G$). Each "transaction" $\Phi^k$ has its number $q$. This number defines the probability $P(q)$ of the element $\Phi_q^k$ in the database $DB_0$: $P(q) \equiv p(\Phi_q^k) = f(\Phi_q^k)/N$. A one-dimensional array $Sign(q)$ is filled by ones. If in the process of analysis it appears that $\Phi_q^k$ is a reducible FP, than the operation $Sign(q) = 0$ will be performed.

We also investigated a possibility of selection of IFPs which overcome the threshold (10) in the process of directly applying the main algorithm, without generating of the manifold $G$. However, this method complicates the logic scheme of the procedure and requires much more memory, and does not reduce the searching time. That is why, for searching IPFs we used the procedure which includes a generation of the manifold $G$.

**Exclusion of reducible FPs from the manifold $G$.** This procedure is logically rather simple, but in order to reduce the computational time it requires a specific realization for databases with "long-range" interactions (IPFs with long length).

The candidates in IFPs that belong to the manifold $G$ are denoted as $\Phi_j^k$. The upper index indicates the length of the FP, the lower index indicates the number in the general list of elements in G (see Fig. 10). If $\Phi_j^k$ can be represented as a product $\Phi_m^l \Phi_n^{k-l}$, than one needs to find the level of correlations between $\Phi_m^l$ and $\Phi_n^{k-l}$. If it appears to be smaller than 1, than $\Phi_j^k$ should be removed from $G$.

The operation of testing on belonging to the set of IFPs we perform consistently for the groups of elements $\Phi^l$ with the same length starting with $l = 3$. Suppose that all $\Phi^{l<k}$ were tested on belonging to IFPs, and than we start the analysis of the group of elements $\Phi^k$. We demonstrate the procedure of this testing on the following example (see Fig. 10).

1. Choose the element $\Phi_m^3$ in the group with elements of the length $l = 3$ (we assume that a described below procedure with the elements $\Phi$ located to the left from $\Phi_m^3$ (see Fig. 10), we already performed.

2. Search in the group of elements of the length $k$ the element $\Phi_j^k$, which includes the element $\Phi_m^3$.

3. Define the element $\Phi_{j,m}^{k-3}$ (quotient) of the length $k-3$, which supplements $\Phi_m^3$ to $\Phi_j^k$. Formally, this procedure can be represented as follows: $\Phi_{j,m}^{k-3} = \Phi_j^k / \Phi_m^3$.

4. Find out if FP $\Phi_{j,m}^{k-3}$ is in the manifold $G$ in the group of elements of the length $k-3$. If such an element (and also its number, $n(j,m)$) exists, we verify the criterion (10), represented in the form

$$\left| \frac{P(j)}{P(m)P(n)} - 1 \right| > \gamma \sqrt{\frac{1 - P(m)P(n)}{N \cdot P(m)P(n)}} . \qquad (21)$$

5. If the criterion (10) is satisfied, than we continue the analysis of the group of elements with the length $l = k$ in order to find the next element $\Phi_{j'}^k$, which includes $\Phi_m^3$. If the criterion (10) is not satisfied, than we exclude $\Phi_j^k$ from the list of IFPs, changing the value of $Sign(j)$ from 1 to 0: $Sign(j) = 0$.

6. After completing the operation with $\Phi_m^3$ we pass to the described procedure with $\Phi_{m+1}^3$, and so on, ending the procedure with the last element in the group with the length $k-1$.

7. In the general case, the ratio $\Phi_{j,m}^{k-p} = \Phi_j^k / \Phi_m^p$ could not be contained in the manifold $G$. In this case, we should refer to the database $DB_0$, in order to find in it the frequency $f(\Phi_{i,m}^{k-p})$ to verify the criterion (21), in which one should use $P(n) \equiv f(\Phi_{j,m}^{k-p})/N$. (One cannot build the frequency database $f(\Phi_{j,m}^{k-p})$ on the stage of generation of the manifold $G$, as one does not know in advance the total set of $\Phi_{j,m}^{k-p}$).

For large maximum lengths κ of the selected candidates in IFPs, it is necessary to systemize the performed operations in order to avoid redundancies. Below we present one of the variants of such systematization.

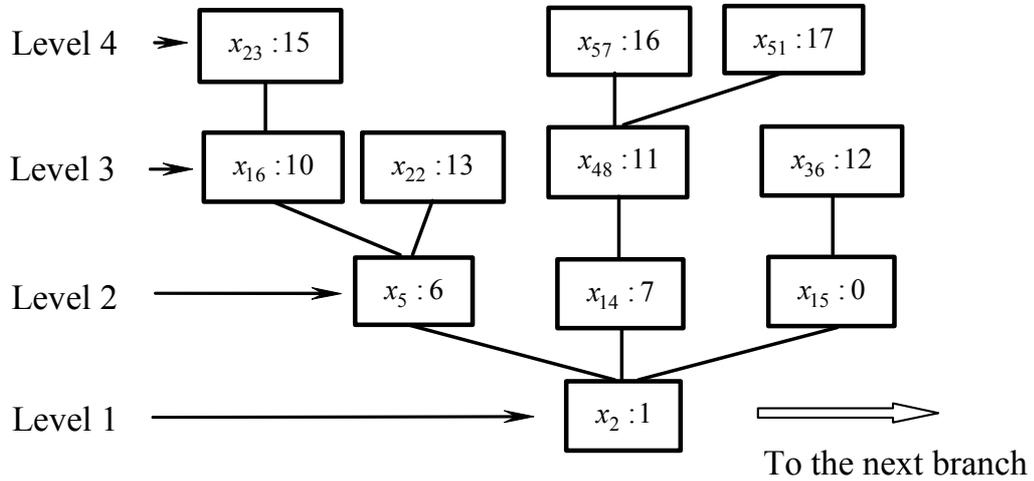

Fig. 11. The example of one branch of the tree $T^4$ for elements $\Phi^{l \leq 4}$ of $G$ starting with $x_2$. One starts from the beginning of the branch and, moving along the links, finishes in the arbitrary node of the tree. The elements of all nodes of the passed chain form FP, $\Phi \in G$. The number in the final node ($q$) indicates the corresponding number of FP in G.

Suppose, that we start testing the group of elements $\Phi^k$ with the length $k$. In each element of this group we should find more shorter elements $\Phi^{l<k}$. Then, we find the quotients $\Phi^{k-l}$, and verify the criterion (21). To fulfill this procedure, the following steps are made.

**A**. For the set of all elements $\Phi^{l<k}$ we build a tree $T^{k-1}$ in a standard way (see, for example, [5,6]). Each branch of this tree includes all elements $\Phi^{l<k}$, starting with the element $x_j$ of group $\Phi^1$. Fig. 11 demonstrates an example of building the branch of such tree for all elements starting with $x_2$. If to choose the element on this branch on the level $n$ and, starting with it, come back from the top to the bottom to the beginning of the tree (or to complete a trip in an opposite direction), than the corresponding chain of elements $x_{i_n} x_{i_{n-1}} ... x_{i_1}$ represents an element (column) of the manifold $G$ with some ordering number $q$ (see Fig. 10). These numbers $q$ are indicated for the nodes of the branch of the tree in Fig. 11. For example, if to start with the element $x_{23}$ of the level 4, than we get the element $\Phi^4_{15} = x_{23}x_{16}x_5 x_2$ located in $15^{th}$ column of $G$. If at the name of the element index 0 appears, than the corresponding element does not exist in the set $\Phi^{l<k}$. For example, $x_{15}$ on the second level is included in $\Phi^3_{12} = x_2 x_{15} x_{36}$, but the pair $\Phi^2 = x_2 x_{15}$ is not included in $G$. The whole tree for elements $\Phi^{l<k}$ in $G$ represents an operating system for the step $B$.

**B**. On this step, we consider a manifold $\Phi^k$ as some database and search for FPs in it using the main algorithm described in [15], with a threshold $\xi = 1$. Below, we shall denote this procedure as $P_{fp}(\Phi^k)$. The purpose of this procedure is in the fast search of the remains $\Phi^{k-l}_{j,m} = \Phi^k_j / \Phi^l_m$, $l < k$. When this procedure is performed, the remains $\Phi^{k-l}_{j,m}$ are represented by the conditional databases which are generated on the level $l$ for radicals $\Phi^l_m$. Note, that when processing the manifold $\Phi^k$, we do not need to build all FPs (radicals of the level $l$) which are contained in it, but only those which represent the elements $\Phi^{l<k}$ in $G$. Just for this purpose, we shall use the tree built in A, as an operating system of $P_{fp}(\Phi^k)$ for optimization of the computational algorithm. Remind, that in the algorithm [15] the database $DB_0$ and all conditional databases $DB_l$, which are designed on the levels $l$, are represented by the lists of the numbers of transactions which include the radicals $u^l = x_{i_1}, x_{i_2}, ..., x_{i_l}$. These numbers are settled in the database $DB_0$ before the computational process starts. Namely, by calculating the frequency of any FP in $DB_0$ we simultaneously define the numbers of transactions in which this FP appears. As will be seen below, this is very convenient when analyzing the manifold $G$.

How is the tree $T^{k-1}$ used while performing the procedure $P_{fp}(\Phi^k)$? For example (see Fig.11), the elements $x_2$, $x_5$ are connected by links, and then the elements $x_{16}$ and $x_{22}$ are connected with $x_5$. This means that first in the database $\Phi^k$ the transactions with numbers $j(x_2)$ will be found and studied on belonging to the set of IFPs which include the element $x_2$. Then, the operating tree directs us to the search and analysis of transactions $\Phi^k$ with numbers $j(x_2 x_5)$, which include the pair $x_2 x_5$. These numbers, $j(x_2 x_5)$, we should search not in the whole set of transactions but only in the known at this point the set $j(x_2)$. Analogously, $j(x_2 x_5 x_{16})$ and $j(x_2 x_5 x_{22})$, according to Fig. 11, will be searched only in $j(x_2 x_5)$. Actually, this means that the elements $x_{16}$ and $x_{22}$ will be searched on the level 3 of the procedure $P_{fp}(\Phi^k)$ in the conditional database produced from $\Phi^k$ on the level 2 of $P_{fp}(\Phi^k)$ for the radical $x_2 x_5$. The use of the tree allows us to utilize effectively the results of the previous search for the new one, which significantly reduces the computational time. Thus, the beginning of each branch of the tree $T^{k-1}$ defines the basic elements (radicals) on the level 1 of $P_{fp}(\Phi^k)$. The elements of the level 2 of each branch of the tree $T^{k-1}$ (Fig. 11), which are connected by links with elements of level 1, define the basic elements of level 2 of $P_{fp}(\Phi^k)$, for a given element of level 1, and so on. Thus, on the level $Lev$ of $P_{fp}(\Phi^k)$ only the elements $\Phi^{l<k}$, $l = Lev$, are generated.

We mention now the data which we possess on each level $Lev$ of $P_{fp}(\Phi^k)$.

1. The elements of a FP in $\Phi^k$, which coincide with elements of $\Phi_m^{Lev<k}$, and also the probability $P(m)$ of a FP in $DB_0$ (the value $m$ is indicated on the tree near the last (top) element $x_{i_{Lev}}$ in $\Phi_m^{Lev<k}$).

2. We know all numbers $j(\Phi_m^{Lev<k})$ of transactions $\Phi_j^k$, which include a given FP, $\Phi_m^{Lev<k}$, as a procedure $P_{fp}(\Phi^k)$ operates on each level with the numbers of transactions defined for $\Phi^k$ in $G$ (see. Fig. 10). Correspondingly, we know the probabilities $p(\Phi_j^k) \equiv P(j)$.

3. And, finally, which is the most important, there is no needs to find the supplements, $\Phi_{j,m}^{k-Lev} = \Phi_j^k / \Phi_m^{Lev}$, as these supplements are the "transactions" which belong to already known conditional database $(\Phi^k)_{Lev}$, derived on the level $Lev$ for the radical $\Phi_m^{Lev<k}$.

To verify the belonging of $\Phi_j^k$ to the set of IFPs we need to know the probability $p(\Phi_{j,m}^{k-Lev})$ in $DB_0$. For this, we will find the element $\Phi_{j,m}^{k-Lev}$ in $\Phi^{l<k}$ using again the tree $T^{k-1}$, which was built in A*. If such an element, $\Phi_{j,m}^{k-Lev}$, exists in $\Phi^{l<k}$ (is found on the tree $T^{k-1}$), then its current number, $n(j,m)$, will be determined in $G$. Thus, $p(\Phi_{j,m}^{k-Lev}) \equiv P(n)$. Then, we verify the criterion (21), by using $P(j)$, $P(m)$ and $P(n)$. In relation to this verification we either leave (equal to 1) or change to 0 the value of B $Sign(j)$. If the results of this search are negative, then we put the remainder, $\Phi_{j,m}^{k-Lev}$, in the subsidiary database $Rest$. The structure of this database is shown in Fig. 12. (The analysis of the database $Rest$ will be completed on the final stage of the analysis of the manifold $G$.)

---

* We will not describe here additional properties of the tree for an accelerating search due to their wide discussion in the literature.

| $x_{12}$ | $x_{65}$ | | "Transactions" of the database of the unknown components | | | | | | | $x_{89}$ $x_{55}$ • • • $x_{39}$ |
| $x_2$ | $x_{57}$ $x_{14}$ $x_2$ | | | | | | | | | |
| 1 | 2 | 3 | ••• Number $n$ of the $\Phi_{j,m}^{k-l}$ in $Rest$ | | | | | | | |
| Number $j = J(n)$ of the "parents" $\Phi_j^k \in G$ | | | | | | | | | | |
| Number $m = M(n)$ of the known components $\Phi_m^{l<k} \in G$ | | | | | | | | | | |

Fig. 12. The structure of the database $Rest$ of the remainders, which were not found in the manifold $G$. Two bottom one-dimensional arrays $J(n)$ and $M(n)$ keep the information about the numbers $j$ and $m$ of elements of $G$, which provided the remainder $\Phi_{j,m}^{k-Lev}$; $n$ is the ordering number of $\Phi_{j,m}^{k-Lev}$ (column) in the database $Rest$.

After the end of the procedure $P_{fp}(\Phi^k)$, we build the operating tree $T^k$, by adding to the tree $T^{k-1}$ "transactions" of $\Phi^k$. Then, we move to the procedure $P_{fp}(\Phi^{k+1})$. Note, that the tree $T^k$ includes in itself the unchanged tree $T^{k-1}$, and also includes new branches and becomes longer. The last performed procedure is $P_{fp}(\Phi^L)$.

Finally, we move to the step C.

**C.** We need to test on "irreducible" property the elements of $G$ relatively to the remainders in the database $Rest$. The database $Rest$ is not as big as it could look. In our numerical simulations with artificial databases, its size is a few times smaller than the size of $G$ in both (i) the number of transactions, and (ii) their lengths. Note, that if the element $\Phi_j^k \in G$ produced the remainder ("transaction" in $Rest$) under the test of its elements $\Phi_m^{l<k}$, than the same $\Phi_j^k$ could not pass later a test by other elements $\Phi_{m'}^{l'<k}$. This happens rather often. Thus, it is necessary to organize a "cleaning" of the database $Rest$ from unwanted "transactions", after performing each of procedures $P_{fp}(\Phi^k)$, $3 \leq k \leq L$. A sign of removal of the remainder with the current number $n$ in $Rest$ is the condition $Sign[J(n)] = 0$.

To finish the test of the manifold $G$ we should find the frequencies (probabilities) of the remainders, $\Phi_{j,m}^{k-Lev}$, in the database $DB_0$. To do this, we perform the procedure which is analogous to one described in A and B. For transactions of the database $Rest$ we build a tree which will be operating when analyzing the database $DB_0$. When processing the database $DB_0$, we shall find the probabilities $P(n) \equiv p(\Phi_{j,m}^{k-Lev}) = f(\Phi_{j,m}^{k-Lev})/N$. Besides, the variable $n$ points on the values $P(m) \equiv P[M(n)]$ and $P(j) \equiv P[J(n)]$ in criterion (21) for testing the element $\Phi_j$ in $G$. The last step is to collect irreducible elements $\Phi \in G$. To do this, we need overlook a one-dimensional manifold $Sign(j)$ and extract only those $\Phi_j$, for which $Sign(j) = 1$.

A described procedure of testing the elements of $G$ requires the computational time not larger than the process of formation of this manifold. The total computer time required for selection of IFPs in our numerical experiments with artificial databases $DB_0$ (even for small threshold $w_0$ ($w_0 = 0.25$)) was by, at least, one order of magnitude smaller than the time needed for searching all FPs in $DB_0$. Under the "express-analysis" of $DB_0$ ($w_0 = 1$) the gain in computer time reaches up to two orders of magnitude.

| Length $l$ of elements $\Phi \in G$ | Number of elements $\Phi$ of length $l$ in $G$ | Number of elements $\Phi$ of length $l$, passed test on irreducible patterns on the step $B$ | Number of IFPs of length $l$ selected on the step $C$ |
|---|---|---|---|
| 1 | **90** | **90** | **90** |
| 2 | **316** | **316** | **316** |
| 3 | **3.126**   (891) | **895**   (778) | **800**   (727) |
| 4 | **12.518**   (1019) | **3.573**   (735) | **704**   (555) |
| 5 | **24.368**   (540) | **6.697**   (324) | **286**   (177) |
| 6 | **20.846**   (145) | **4.986**   (74) | **55**   (28) |
| 7 | **8.007**   (8) | **1.811**   (3) | **4**   (0) |

Table 2. Dynamics of the number of candidates in IFPs in the process of testing elements $\Phi^l$ of $G$ for belonging to IFPs: $w_0 = 0.3$, $R = 4$. In parentheses the analogous results are shown for $w_0 = 1$.

The Table 2 demonstrates the testing results of the manifold $G$ on the steps $B$ and $C$. A visible difference in the results of the "express-analysis" and of more exact consideration ($w_0 = 0.3$, $R = 4$) reveals for IFPs of the length $l = 5$ and more. The use of parameter $R$, which limits the order of registered interactions, can be rather useful for analysis of the database $DB_0$. For example, a decrease of this parameter from $R = 4$ to $R = 3$ reduced the number of elements in $G$ by factor 1.5. However, the final result after testing this manifold did not change according to Table 2 (only the numbers of IFPs of the length 5 and 6 were reduced by 1). One can conclude, that in the database $DB_0$ the interactions of the third and of more higher orders do not exist. In the database with weak interactions (correlation coefficients were generated in accordance with the rule (13) for $\theta = 0.2$, $p_{max} = 0.6$) the number of IFPs is reduced and equals to 197, 234, 107, 19, 1 and 0 for lengths $l = 2, 3, 4, 5, 6, 7$, correspondently ($w_0 = 0.3$, $R = 3$).

When analyzing a database with an unknown structure, it is useful to undertake first an "express-analysis" and 2-3 additional variants by varying the parameters $w_0$ and $R$ (total computational time required for these procedures is still smaller than the time required for searching all FP). The dynamics of the results obtained will prompt a strategy of the required final computational procedure. We are convinced that in the database of the arbitrary physical nature one needs to select only IFPs, the number of which is significantly smaller than the number of transactions, and which include all important information about the structure of the database and the interactions of its elements.

## 3. Effectiveness of IFPs

To illustrate the effectiveness of the IFPs approach in application to classification problem we consider here two databases, A and B, with the transactions generated according to some rules. The problem is (i) to recognize the rules which were used to generate the transactions in A and B and (ii) make a decision on belonging of a new transaction to A or B. We argue that just IFPs represent those rules which allow one to solve this problem.

**Method for recognition of transactions.** In each database, A and B, the particular interactions between the elements and the particular sets of IFPs exist: $G_A$ and $G_B$. Each transaction $T$ includes patterns which belong to both $G_A$ and $G_B$. These patterns represent the sub-sets, $g_A \in G_A$ and $g_B \in G_B$. Generally, $g_A \neq g_B$. Introduce the sum, $g_{AB} = g_A \cup g_B$. The elements $g_{AB}$ will be denoted as $\Phi_i^l$, where $l$ is the length of the pattern, and $i$ is its number in $g_{AB}$. The product $g_{ab} = g_A \cap g_B$ includes patterns which are IFPs in both databases A and B. All other patterns $\Phi_i^l$ are the IFP in one database, but are not IFPs in the other one. We will denote the frequencies of patterns $\Phi_i^l$ in A and B as $f_{i,a}^l$ и $f_{i,b}^l$ (for simplicity of presentation and to avoid the normalization issue we assume that the numbers of transactions in A and B are the same). Any transaction $T$ will be characterized by the scoring factor [16]

$$S(T) = \sum_{l,i} \ln\left(f_{i,a}^l / f_{i,b}^l\right). \tag{22}$$

The leaning procedure for pattern recognition is performed on the initial databases A and B. For each transaction $T_a$ of the database A we calculate $S(T_a)$ and build the distribution

$$n_a(S) = N_a(S)/\Delta S, \tag{23}$$

where $N_a(S)$ is the number of transactions in A with the scoring factor in the interval $(S - \Delta S/2, S + \Delta S/2)$. Analogously, we calculate the dependence $n_b(S)$ for B. As a rule, the dependences $n_a(S)$ and $n_b(S)$ have the bell structure with maxima which are shifted one from the other along the $S$ axis. The accessory of the new transaction $T$ to A or B is established by the parameter $\vartheta = n_a(S(T))/n_b(S(T))$. If $\vartheta > 1$ (or $\vartheta < 1$), than it is more probable that the transaction $T$ belongs A (B). For the intermediate cases one can prescribe the transaction $T$ to A (B) with the probabilities

$$\begin{aligned} P(T \in A) &= n_a(S)/(n_a(S) + n_b(S)) = \vartheta/(\vartheta+1), \\ P(T \in B) &= 1/(\vartheta+1). \end{aligned} \tag{24}$$

**Generation of the artificial databases.** The transactions of A and B are built by using 30 variables, $Y_k$. When generating these databases only the interactions of the third order are introduced, and all other interactions appear in an uncontrolled way. As in Section 2, each variable can take only one of three values: $y_{km}$, $k = 1, 2, 3, \ldots\ldots 30$, where $m = 1, 2, 3$. Instead of variables $y_{km}$ we use the variables $x_l$, $l = 1, 2, 3, \ldots\ldots 90$, where $l = 3(k-1) + m$. The number of transactions in the databases is $N = 200000$. A generation of the value of the variable $x_l$ in transactions is done as follows [remind that $x_l$ can take two values: $x_l = 0$ (absent in a given transaction) and 1 (present in a given transaction)].

The probability $p(x_l) = p_{\max} \times \zeta$ ($0 \leq \zeta \leq 1$ of $x_l = 1$ in a transaction is chosen randomly, where $p_{\max}$ is a parameter). Introduce the correlations of the third order by using the conditional probabilities

$$p_{11} = p(x_l = 1 | x_i x_j = 1) = c_{ij,l} \cdot p(x_l),$$
$$p_{10} = p(x_l = 1 | x_i x_j = 0) = p(x_l)[1 - c_{ij,l} \cdot p(x_i x_j)] / [1 - p(x_i x_j)], \qquad (25)$$
$$c_{ij,l} = 1.9 + \theta \times \zeta.$$

Here $\theta$ is a parameter, $p(x_i x_j)$ is a ratio of the frequency $f(x_i x_j)$ to the number of transactions $N$; indices $i, j < l$, which identify the interacting elements, are taken randomly. The following constrain should be used: $((l-i), (l-j) > 2, |i-j| > 2)$. The value $x_l$ is taken by the following rules. Choose the random number $\zeta$.

$$\text{If } x_i x_j = 1, \text{ than } x_l = \begin{cases} 1, & \zeta \leq p_{11} \\ 0, & \zeta > p_{11} \end{cases}; \text{ if } x_i x_j = 0, \text{ than } x_l = \begin{cases} 1, & \zeta \leq p_{01} \\ 0, & \zeta > p_{01} \end{cases}.$$

On average over all transactions in the database, the frequency of the event $x_l = 1$ is $f(x_l = 1) \approx N \cdot p(x_l)$. Generation of the first six values of variables ($l \leq 6$) is done without imposing correlations between them.

When generating the databases A and B the parameter $p_{\max}$ was chosen $0.6$. The parameter $\theta = 0.1$ was chosen for the database A. For the database B all $c_{ij,l}$ were chosen 1 (absence of correlations), correspondingly in (25) $p_{11} = p_{10} = p(x_l)$. Thus, in the illustrative example we have the database A with correlations and the IFPs set $G_A$, and the database B without correlations. Note, that the frequencies of elements in A and B are practically the same and, consequently, do not provide additional information for recognition of transactions with given rules.

**Results of numerical simulations.** For the selection parameter $\gamma = 2$ (see (11)) of the candidates into the IFPs for the database A, the number of IFPs of the length $l = 2, 3, 4, 5, 6, 7$ is $f(l) = 676, 3193, 6099, 4283, 1248, 167$, correspondingly. Fig. 13 demonstrates the dependences $n_a(S)$ and $n_b(S)$. The vertical lines indicate the boundaries of the fuzzy region for recognition of transactions by using a scoring factor. On the left boundary $n_b / n_a = 2$, and on the right boundary $n_a/n_b = 2$. Inside the region $F$ the probabilities $P(T \in A)$ and $P(T \in B)$ differ not more than by factor 2. For the considered databases approximately a half of all transactions belong to the fuzzy region: $N_f \approx N/2$. The parameter $R = N_f/N \approx 1/2$ can be used as a measure of error provided by this method. In the considered case of databases the value of this parameter is not small enough. But it was not our intension here to choose the databases which are characterized by a sharp boundaries between the curves $n_a(S)$ and $n_b(S)$. Our goal is to elucidate the significant role of IFPs in the recognition of new transactions.

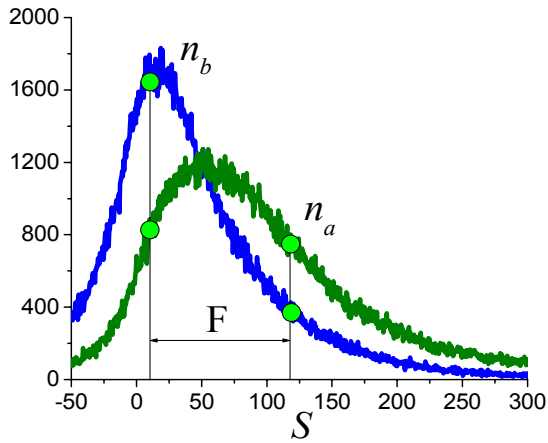
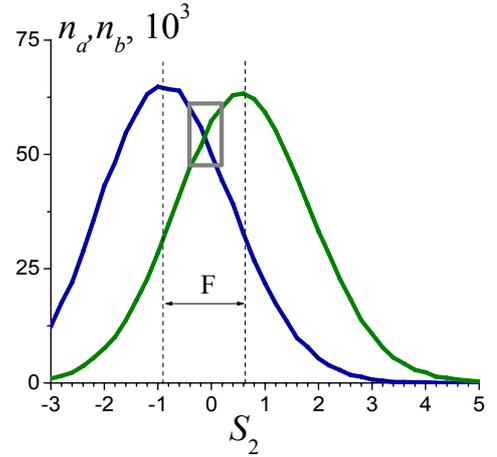

Рис. 13. Dependences $n(S)$ for the database with correlations ($n_a$) and without correlations ($n_b$). 110 thousand transactions from each database appeared in the fuzzy region F.

Рис. 14. Scoring (recognition) curves for the partial scoring factor $S_2$. 75000 transactions from the database A and ~ 82000 transactions from B belong to the region F.

**The following conclusions can be made from the numerical experiments:**

**A.** The information provided by the IFPs about the databases is distributed inhomogeneously throughout the IFPs with a given length $l$. Introduce the partial scoring factor $S_l$:

$$S_l(T) = \sum_i \ln\left(f_{i,a}^l / f_{i,b}^l\right), \qquad (26)$$

which, opposite to (22), includes only the IFPs of length $l$ from the set $g_{ab}$. (Remind that for each transaction a different set $g_{ab}$ exists.) Fig. 14 demonstrates the dependences $n_a$ and $n_b$ ("scoring curves" or "recognition curves") for the partial scoring factor $S_2$. The parameter $R$ decreases a little for these curves (at least, it did not increase in comparison with the results shown in Fig. 13). This indicates a possibility to characterize (and recognize) transactions using only a partial scoring factor with $\gamma = 2$. On the other hand, the IFPs of the length $l = 3$ often are built from IFPs of the length $l = 2$ by adding the third element. That is why, one can expect that the information about the databases provided by the IFPs $\Phi_i^3$ is more complete than one provided by $\Phi_i^2$. It is difficult to make a general conclusion at this time, but in the considered example of the databases the use of the scoring factor $S_3$ (see Fig. 15) does not improve the quality of recognition of transactions in comparison with the results shown in Fig. 14 (does not reduce the value of parameter $R$), but even makes it worse.

**B.** A successive application of the partial scoring factors can improve the quality of the recognition procedure. The effectiveness of this procedure depends on the order in which the partial factors are applied. Explain this by an example. In Fig. 14, in the gray rectangular the transactions are indicated

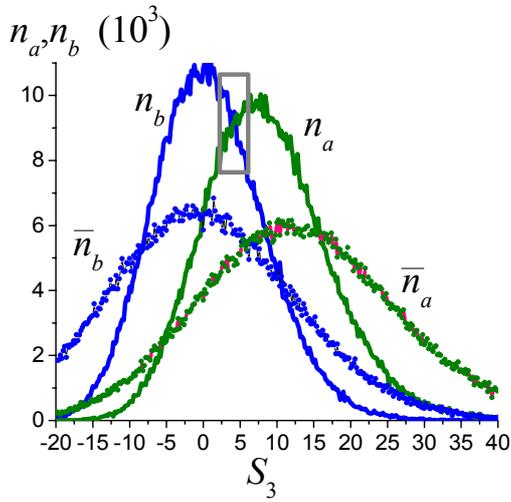

Рис. 15. The scoring curves built by using a partial factor $S_3$ (in fuzzy region $n_a$ and $n_b$ appear approximately 103 thousand transactions from each database). The dependences $\bar{n}_a$ and $\bar{n}_b$ are

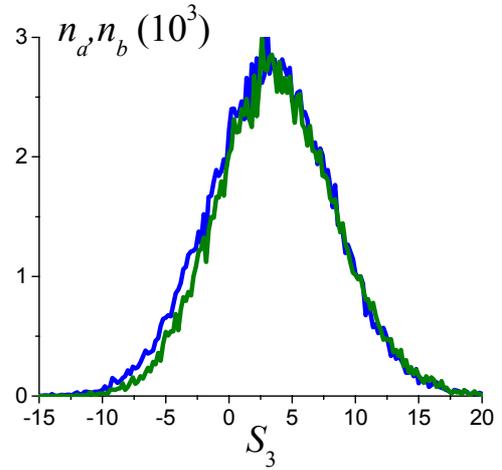

Рис. 16. Scoring curves for transactions which belong to the region with high uncertainty (gray rectangular in Fig 14), built by using a partial scoring factor $S_3$.

(their number is approximately 33 thousand in each database) with high level of uncertainty identification. We shall build for these transactions the scoring curves by using the partial scoring factor $S_3$. Namely, we use subsequently two "filters" to identify the transactions. First, we use a scoring $S_2$, and then the scoring $S_3$. The second filter appeared to be useless in this case (see Fig. 16). Thus, the results presented in Fig. 16 indicate the low level of effectiveness of application for *chosen databases* the scoring procedure by using the scoring factor $S_3$ after $S_2$. The effectiveness of IFPs of the length $l$ for identification of transactions in the database $DB_0$ (independently of the method for calculation of the scoring factor) can be determined by the parameter $U_l = \bar{n}_{DB}(\Phi^l)/N_{DB}(\Phi^l)$, where $N_{DB}(\Phi^l)$ is the total number of IFPs of length $l$ in the database $DB_0$, and $\bar{n}_{DB}(\Phi^l)$ is the average number of such IFPs in transactions in the database $DB_0$. In our case, in the database A $U_3 = 114/3193 < U_2 = 45/676$. The parameter $U_l$ characterizes the "activity" of IFPs of the length $l$ in identification of transactions.

Now we use the opposite sequence of operations. Select in Fig. 15 (scoring curves for $S_3$) similar transactions with high uncertainty (gray rectangular, with the number of these transactions approximately, as before, 33 thousand for each database) and put them though the second filter using the partial scoring $S_2$. Fig. 17 demonstrates the effectiveness of using two filters in this case: for many transactions which were initially the ratio of the probabilities $P(T \in A)$ and $P(T \in B)$ exceeds 2. If one applies the partial scoring $S_2$ to all transactions in the fuzzy region in Fig. 15, than after two filters $S_3$ and $S_2$ (Fig. 17) the possibilities to identify the transactions improve relatively to the case of applying only the partial scoring $S_2$. In the fuzzy region appear about 62-64 thousand transactions from each database A and B.

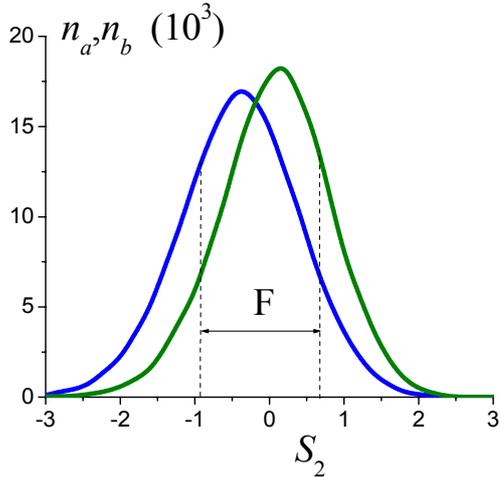 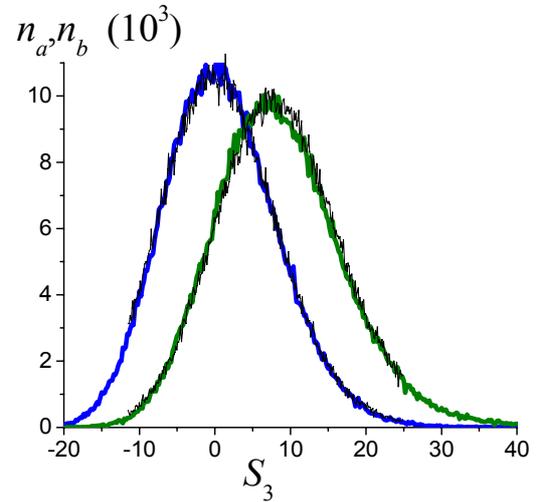

Fig.17. Scoring curves for transactions, which belong to the region with high uncertainty (gray rectangular in Fig 15), built by using a partial scoring factor $S_2$.

Fig. 18. A demonstration of the similarity of scoring curves for $\gamma = 2$ (curves $n_a, n_b$) and $\gamma = 1$ (curves $\bar{n}_a, \bar{n}_b$) in Fig. 15. The transformed dependences $\bar{n}_a, \bar{n}_b$ are shown by black curves.

**C.** Calculation of the scoring factor by using all FPs is rather unnecessary procedure, because the information about the databases is contained mainly in the IFPs. We carried out the following numerical experiment. We decreased the threshold for selection of candidates to IFP from $\gamma = 2$ to $\gamma = 1$. Moreover, in the generated set of candidates (this set was called $G$ in Section 2) a selection of IFP was not done. As a result, the number $\Phi^l$ of different length increased significantly. For example, for $l = 2, 3, 4$ the corresponding frequencies are $f(l) =$ 836, 10783, 49508. At the same time, the scoring curves built on the partial scoring factor $S_2$ did not change. The results of using the partial scoring factor $S_3$ for different values of $\gamma$ are shown in Fig.

15. For comparison of the effectiveness of the scoring curves $n_a, n_b$ and $\bar{n}_a, \bar{n}_b$ in Fig. 15 we transformed the function $\bar{n}_a(S_3)$ into the function $S_3$ by using the linear transformations. As a result of these transformations, the function $\bar{n}_b(S_3)$ transformed into the function $n_b(S)$, see Fig. 18. This shows that no changing of the accuracy of identification (its improvement) appears. An analogous result was obtained for the partial scoring factor $S_4$, and also for the case when the integrated scoring was calculated for $2 \leq l \leq 4$. Thus, the increasing of the number of FPs, which are used in building a scoring factor, up to 63 thousand does not improve the recognition parameter $R$, which was calculated for 10 thousand IFPs. The obtained results illustrate the above made conclusion: generation of a total number of FPs is unnecessary procedure for the practical use of the rules of the database in the problem of identification of transactions; these rules are represented by much smaller number of IFPs.

## 4. Conclusion

**Relation to the Market Basket Analysis (MBA).** In the MBA a given transaction can be characterizes by only IFPs of different lengths. All FPs which are built from the non-interacting elements $x_i$ do not provide new information.

**Relation to the scoring of transactional databases.** When building a scoring factor for transactional databases the IFPs, which correspond to a given transaction, can be included from "Good" and "Bad" databases with different weights. Or one can use as a weight the parameter $|\chi| = \left| \frac{f(v)}{Nq_{ux}} - 1 \right|$, which is calculated in the main algorithm for each IFP before a selection procedure. This parameter (a left side in criterion (10)) characterizes the degree of reliability that a given FP is built from the interacting elements and belongs to the set of IFPs. Another possibility, which was utilized above, is to use as a scoring factor, $s$, for a transaction a function of the type $s = \sum_n [p_{good}(u_n) / p_{bad}(u_n)]$, in which only the IFPs $u_n$ should be included.

**Relation to the bio-systems.** In bio-systems a complex molecular can usually be encoded by similar variables which are used in transactional databases. Then, the statistical analysis of an ensemble of such molecules is reduced to the analysis of the transactional databases. The number of FPs can be very large in this case. At the same time, the number of IFPs can be many orders of magnitude smaller than the number of IFPs. These IPFs can serve as characteristic features (rules) of the considered bio-system.

**Relation to the immune systems and the vaccination problem**. In the immune systems the structures of viruses can be coded in a similar way as transactions in transactional databases, by a set of digital variables, $x_i$. In this case, only the IFPs represent new information about the viruses, and all FPs with non-interacting elements can be neglected. If statistical approach is assumed for solving a vaccination problem, it looks possible that a relatively small number of vaccines needed to prevent many probable diseases can be related to a small number of IFPs which will appear in the virus representation. Indeed, only a "struggle" with IFPs of viruses is needed. This could explain why a relatively small number of vaccines are needed to struggle with much larger number of viruses.

In conclusion, we note that the method of selection of IFPs is applied to the random values—the frequencies of the patterns of different lengths. That is why, the problem of belonging of a given FP to the set of IFPs is of the probabilistic nature, and includes some uncertainty. But, in any case, a set of IFPs more definitely characterizes a given database than a much larger set of FPs, the majority of which are built from the non-interacting elements. There are three main advantages for using the IFPs: (i) each IFP represents, up to some extent, new information about the database, (ii) the number of IFPs is significantly smaller than a total number of FPs, and (iii) the search for IFPs is significantly more efficient than the search for FPs. These features of IFPs are very important not only for characterizing a database but also for classification of the new transactions and making decisions about new transactions. Besides, the operations with tens and hundreds of millions of FPs are very time-consuming and, in principle, limit the analysis of large databases and classification of new transactions. The random frequencies of many non-representative FPs which are built from non-interacting elements could create a very high level of noise and prevent to separate useful signals which characterize database and new transactions.

**Acknowledgments**

This work was supported by the Department of Energy under the contract W-7405-ENG-36 and DOE Office of Basic Energy Sciences.